\documentclass{article}
\usepackage{arxiv}

\usepackage[utf8]{inputenc} 
\usepackage[T1]{fontenc}    
\usepackage[bookmarks=false]{hyperref}       
\usepackage{url}            
\usepackage{booktabs}       
\usepackage{amsfonts}       
\usepackage{nicefrac}       
\usepackage{microtype}      
\usepackage{lipsum}
\usepackage{cite}
\usepackage{graphicx}
\usepackage{amssymb}
\usepackage{amsmath}

\title{A Hybrid Deep Learning Model for Predictive Flood Warning and Situation Awareness using Channel Network Sensors Data}

\author{
  Shangjia Dong \\
  Department of Civil and \\
  Environmental Engineering\\
  University of Delaware\\
  Newark, DE 19711\\
  \texttt{sjdong@udel.edu}\\
   \And
 Tianbo Yu \\
 Department of Electrical and \\
 Computer Engineering\\
  Texas A\&M University\\
  College Station, TX 77840\\
  \texttt{tianbo@tamu.edu}\\
  \And
  Hamed Farahmand \\
  Zachry Department of Civil and \\
  Environmental Engineering\\
  Texas A\&M University\\
  College Station, TX 77840\\
  \texttt{hamedfarahmand@tamu.edu}\\
  \And
  Ali Mostafavi \\
  Zachry Department of Civil and \\
  Environmental Engineering\\
  Texas A\&M University\\
  College Station, TX 77840\\
  \texttt{amostafavi@civil.tamu.edu}\\
}

\begin{document}
\maketitle

\begin{abstract}
The objective of this study is to create and test a hybrid deep learning model, FastGRNN-FCN (Fast, Accurate, Stable and Tiny Gated Recurrent Neural Network-Fully Convolutional Network), for urban flood prediction and situation awareness using channel network sensors data. The study used Harris County, Texas as the testbed, and obtained channel sensor data from three historical flood events (e.g., 2016 Tax Day Flood, 2016 Memorial Day flood, and 2017 Hurricane Harvey Flood) for training and validating the hybrid deep learning model. The flood data are divided into a multivariate time series and used as the model input. Each input comprises nine variables, including information of the studied channel sensor and its predecessor and successor sensors in the channel network. Precision-recall curve and F-measure are used to identify the optimal set of model parameters. The optimal model with a weight of 1 and a critical threshold of 0.59 are obtained through one hundred iterations based on examining different weights and thresholds. The test accuracy and F-measure eventually reach 97.8\% and 0.792, respectively. The model is then tested in predicting the 2019 Imelda flood in Houston and the results show an excellent match with the empirical flood. The results show that the model enables accurate prediction of the spatial-temporal flood propagation and recession and provides emergency response officials with a predictive flood warning tool for prioritizing the flood response and resource allocation strategies.
\end{abstract}


\section{Introduction}
\label{sec:introduction}

Flooding is amongst the most disastrous natural hazards and severely impact communities' health and economic well-being \cite{dong2020a}. Between 1980 and 2017, 227 weather disasters hit the U.S. \cite{winfree2019need}. Hurricane Harvey in 2017 is the costliest tropical cyclone and caused 80 deaths. Additionally, over 100,000 homes are estimated to be flooded and 80,000 of which are estimated to have been flooded at least 0.46 m (18 in), and 23,000 of them were inundated to at least 1.5 m (5 ft) based on Federal Emergency Management Agency Survey \cite{FEMA2017}. The flooding risk is further escalated due to climate change \cite{aerts2018}. However, the increasing frequency and intensity of natural hazards is the one of the contributors to the urban flood risk. Another key reason for the urban flooding is the malfunction of the flood control system in facing extreme rainfall as the channels and drainage system fail to discharge the runoff. The overflow will spread across the network through streets and result in cascading failure. The current flooding warning systems such as Harris County Flood Warning System \cite{HCFWS2019} provide monitoring (based on flood stream gauges). Additionally, there are methods such as Bayesian network model that enable vulnerability assessment in urban areas \cite{dong2019b}. However, these systems and models provide limited predictive flood warning capabilities. Therefore, accurate predictive monitoring of flood risk is vital for improving situation awareness and controlling urban flood risk. 

The standard approach for flood risk modeling is through hydraulic and hydrologic (H\&H) models \cite{itoh2018hydraulic}. Using watershed features, atmospheric exchanges, flow processes, and flood scenarios as input, the H\&H modeling can generate a flood map that indicates the flood depth and the extent of the flood. However, existing H\&H models are primarily applicable for a single watershed analysis to inform flood risk reduction and mitigation decisions \cite{scarlett2018influence}. The ability to use H\&H for predictive flood warning and situation awareness as a flooding event unfolds is rather limited. Since it is critical to acquire the flood information in a predictive fashion for emergency response operations, an accurate network failure (i.e., overflow) prediction model is needed to capture the spatial and temporal failure cascading process in order to better protect citizens and infrastructures from the flooding \cite{dong2019b}. Various methods have been proposed to support real-time flood forecast, including the US National Weather service ensemble forecasting \cite{koren1999scale}, fuzzy reasoning method \cite{liong2000advance}, the river flow forecast system \cite{moore1990basin}, neural networks \cite{thirumalaiah1998real, besaw2010advances}, transfer function methods \cite{young2002advances}, functional networks \cite{bruen2005functional}, and generalized likelihood uncertainty estimation (GLUE) \cite{romanowicz2003estimation}. Data-driven flood prediction methods have gained more popularity as they take much less development time and provide acceptable accuracy in flow prediction to support predictive flood warning and situation awareness \cite{govindaraju2000artificial}. For example, Dong, Yu, et al. \cite{dong2019b} developed a Bayesian network model to predict cascading failure risk of the multiple watersheds flood control system. The model provides an accurate prediction of flooding probability to inform vulnerability assessment. However, the Bayesian network model can scale up fast, with a small increase in the network size, the computation demand grows significantly. Thus, statistical and Bayesian inference methods are not ideal for flood prediction on a large network (with a large number of components and sensors to monitor the infrastructure system). Another way to formulate the problem is treating flood prediction as a time series classification problem. The spatial and temporal records of multiple variables comprise a multivariate time series, where each data item contains variables such as rainfall, the water level in the channel, channel cross-sectional area, and its neighbors' flood status. With the advancements in sensor technology, collecting large-scale flood data from stream gauges is becoming more widely used. The flooding status of different flood control infrastructure can be monitored and recorded over time. Using the historical flood time series data, the future status of the channel infrastructure can be predicted using machine learning-based methods.

Neural networks have been proved to achieve good flood prediction performance. For example, Besaw et al. \cite{besaw2010advances} developed two artificial neural networks (i.e., generalized regression neural network (GRNN) and counter-propagation network (CPN)), to forecast streamflow in basins that do not have flood gauges. Chang, Chang, \& Huang \cite{chang2002real} adopted termed real‐time recurrent learning (RTRL) for stream‐flow forecasting using rainfall-runoff data of the Da‐Chia River in Taiwan. Tiwari \& Chatterjee \cite{tiwari2010development} proposed a hybrid wavelet-bootstrap-ANN (WBANN) model to conduct hourly flood forecasting using five years of hourly water level data from five gauging stations in Mahanadi River basin, India. With the advancement of deep learning techniques, the performance of neural networks has been greatly improved. More recently, deep learning models, such as fully convolutional network (FCN) \cite{wang2017time}, multi-scale convolutions neural network (MCNN) \cite{cui2016multi}, residual network (ResNet) \cite{wang2017time}, and deep multilayer perceptrons (MLP) \cite{wang2017time, widiasari2017deep}, gained strong attention due to their capabilities for multivariate time series classification. In particular, recently developed Fast, Accurate, Stable, and Tiny Gated Recurrent Neural Network (FastGRNN) modeling has shown superior performance \cite{kusupati2018fastgrnn} as it addresses the two limitations of Recurrent Neural Network (RNN): (1) inaccurate training, and (2) inefficient prediction. Additionally, the model structures in RNN is not tailored to the characteristics of the predictive flood warning problem. Moreover, FCN has demonstrated its promising performance in time series classification both in a standalone format \cite{wang2017time} or in hybrid models \cite{karim2017lstm, karim2019multivariate}. Although Bayesian network model and traditional ML methods show good performance in failure detection \cite{dong2019b, dong2020c}, their main advantages lie in assessing the network vulnerability and diagnose the cause of the vulnerability while the hybrid deep learning is able to provide accurate and reliable flood prediction with ability to handle high dimensional data for near real-time prediction to inform situational awareness. Moreover, the growing numbers of sensors being installed in the channel network further increase the network complexity and urge a faster and more efficient model for accurate flood prediction. To address the challenge, this study proposes an end-to-end deep learning time series classification model (FastGRNN-FCN) to predict the channel flooding status change using its nearby flood sensors’ historical spatial-temporal data. The proposed FastGRNN-FCN model in this study enables predictive flood warning and situation awareness at different time steps and locations with high accuracy. The proposed model is tested using flood channel sensors data (stream gauges) related to different flooding events in Harris County, TX. The model outcomes provide stakeholders early warning to inform emergency response and protective actions.

The remainder of the paper is organized as follows. Section \ref{sec:lit_rev} reviews the related literature on flood prediction and multivariate time series classification methods. Section \ref{sec:fastgrnn_fcn} presents the architecture of the proposed FastGRNN-FCN model and its execution procedure. Section \ref{sec:experiment} shows the experiment design for the case study in Harris County and presents an overview of the study site and data processing. Section \ref{sec:results} presents the prediction results and model performance evaluation. Section \ref{sec:discussion} discusses the implications of the model parameters and limitations. Finally, section \ref{sec:conclusion} summarizes the model outcomes and its major findings and future research directions. 

\section{Literature Review}
\label{sec:lit_rev}

The use of sensors for registering flood-related data such as rainfall level and water level in flood gauges as well as the advancements in gathering near real-time crowdsourced data reporting flood events creates excellent opportunities for developing machine learning (ML) models for flood prediction \cite{fan2020social}. Moreover, researchers have developed more computationally efficient ML techniques for multivariate time series prediction that enable highly accurate forecasts when sufficient data entry is available. In this section, we briefly review the state-of-the-art flood prediction research studies enabled by applying data-driven techniques as opposed to standard H\&H models. Next, this section provides an overview of several most recent developments in deep learning (DL) techniques for multivariate time series classification that can be used for near real-time forecasting problems such as predictive flood warning and situation awareness. 

\subsection{Machine learning in infrastructure risk management}
\label{sec:ml_risk_management}

Machine learning (ML) and deep learning (DL) techniques have been widely applied to solve structure and infrastructure engineering problems \cite{hung1999machine}. For example, advanced learning approaches such as deep convolutional neural network–long short‐term memory (CNN–LSTM) have been applied to optimize the use of autonomous vehicles \cite{chen2020deep, wangreinforcement2020}. ML has also been used in the field of structural health monitoring in order to analyze vibration response of structures and detection of anomalies extracted from sensors \cite{azimi2020structural, gulgec2020structural, ni2019deep}. Deep learning has also been used to help analyze traffic crash data \cite{zhang2020ensemble}, analyze crack patterns in concrete structures \cite{cha2017deep, okazaki2020applicability}, and conduct reliability analysis on infrastructure networks \cite{nabian2018deep}.

In recent years, ML techniques have been increasingly employed in infrastructure risk management, including optimized infrastructure maintenance strategies detection \cite{yao2020deep}, failures detection in buildings \cite{rafiei2017novel} and infrastructure networks \cite{wang2020unified}, safety of structures and infrastructures monitoring \cite{rafiei2018novel}, and post-disaster damage and loss estimation \cite{pan2019data}. In particular, ML techniques have shown promising results in infrastructure risk assessment. For example, generative adversarial networks (GAN) is successfully adopted for detecting the road damages \cite{maeda2020generative}. A multilayer ML technique has been used to simulate failure in flood protection systems \cite{fascetti2019multiscale}. ML techniques are also used to develop early warning systems that inform infrastructure protection during disaster events. For example, various neural dynamic classification methods are used for the development of earthquake early warning systems \cite{rafiei2017neews}. 

In the case of flood risk assessment, exposure often is associated with the projected human lives and assets that can be potentially in the areas with high flood risk. In this regard, ML assists urban growth and infrastructure development projection to better recognize areas in the high risk of flood due to high exposure. ML models can also help identify the potential land-use change in the future that requires revisiting flood mitigation measures \cite{hosseini2020flash, jaad2020modeling, wagenaar2019evaluating}. For example, Genetic Algorithm Rule-Set Production (GARP) and Quick Unbiased Efficient Statistical Tree (QUEST) are used to map the flood risk considering factors such as population and urban density as well as socio-economic factors \cite{darabi2019urban}. The flood impact assessment also leverages ML methods for estimating the flood impacts such as flood damage, human impacts such as casualties, and flood losses. To do so, flood-related variables such as flood duration, rainfall intensity, flow velocity, characteristics of buildings and infrastructures, as well as socio-economic factors, are embedded in ML models to develop flood damage functions and flood loss function for predicting the extent of damage and loss in different flood scenarios \cite{kreibich2017probabilistic, merz2013multi}. ML techniques have also been widely utilized for flood hazard modeling such as inundation modeling \cite{alipour2020leveraging} and infrastructure disruption detection using crowdsourced data following hurricanes and floods \cite{roymultilabel2020}. 

\subsection{Flood prediction}
\label{sec:flood_prediction}

Flood inundation models are often used to predict the occurrence of flooding or propagation of flood into different sub-areas of cities. Flood inundation models are categorized into three main categories including hydrodynamic models, empirical methods, and simplified conceptual models \cite{teng2017flood}. Hydrodynamic models employ equations derived from physical laws to simulate the flood propagation process and infer related parameters such as water flow volume and water elevations in different areas \cite{al2003real, balekelayi2019graph, neelz2009desktop, itoh2018hydraulic}. For example, flow rates are often projected employing rainfall-runoff and streamflow models \cite{gori2019characterizing, lu2013streamflow}. As opposed to hydrodynamics methods, empirical methods utilize techniques such as measurements, surveys, ground sensors, and satellite images to forecast the areas that will be inundated and assist the decision-making for mapping the flood risk \cite{darabi2020urban}. 

In addition, network-based data-driven methods also show a good performance in flood prediction. For example, Dong, Yu, et al. \cite{dong2019b} used a Bayesian network technique to detect cascading failure and characterize the network vulnerability in multiple watersheds flood control system. Following that, Dong, Yu, et al. \cite{dong2020c} used the spatial-temporal flood sensor data and employed a network approach to infer the flooding probability of each road intersection considering co-located channel-road network interdependencies. Khac-Tien Nguyen \& Hock-Chye Chua \cite{khac2012data} developed an adaptive network-based fuzzy inference system prediction of water level for a 1-5 day time window. Rainfall-runoff modeling of rivers has also been used for forecasting the river flood with piecewise affine systems, which is identified based on a combined unsupervised clustering-linear regression technique \cite{hadid2020data}. These studies show that data-driven methods hold a strong potential for complementing physics-based models to enhance urban flood prediction.

\subsection{Deep learning in multivariate time series prediction}
\label{sec:dl_time_series}

Deep learning has been widely applied in time series sequences classification \cite{fawaz2019deep}. Various approaches have been proposed, including distance-based methods \cite{orsenigo2010combining}, feature-based methods \cite{pei2017multivariate}, Fisher kernel learning \cite{van2011learning}, Symbolic Representation for Multivariate Time Series (SMTS) \cite{baydogan2015learning}, and ensemble methods \cite{bagnall2015time}, to classify the time series data. All these classifiers require extensive feature engineering \cite{karim2017lstm}. Using an ensemble of these models yields a better result, such as collective of transform based ensembles (COTE) \cite{lines2015time}, proportional elastic ensemble (PROP) \cite{bagnall2015time}, and shapelet ensemble (SE) \cite{bagnall2015time}. Since the historical flood data can be treated as a multivariate time series, deep learning shows great potential for predictive flood warning and situation awareness applications. 

Among the existing deep learning methods, some studies show that FCN can have superior performance in classification problems involving time series \cite{wang2017time, zhang2017resilience}. More recently, long short-term memory fully convolutional networks (LSTM-FCNs) has been developed as an augmentation of FCNs, which has been shown to yield very good predictions \cite{karim2017lstm}. LSTM-FCN architecture employs a global average pooling that enables parameter reduction \cite{lin2013network}. The integrated model comprises an LSTM module with a dropout to prevent overfitting and a fully convolutional module that consists of three stacked temporal convolutional blocks \cite{srivastava2014dropout}. Using the LSTM model and sequence-to-sequence learning, Xiang, Yan, \& Demir \cite{xiang2020rainfall} proposed a rainfall-runoff to predict short-term runoff using rainfall time series data.

\subsection{Deep learning for flood prediction}
\label{sec:dl_flood_prediction}

\begin{figure*}[!ht]
    \centering
    \includegraphics[scale=0.75]{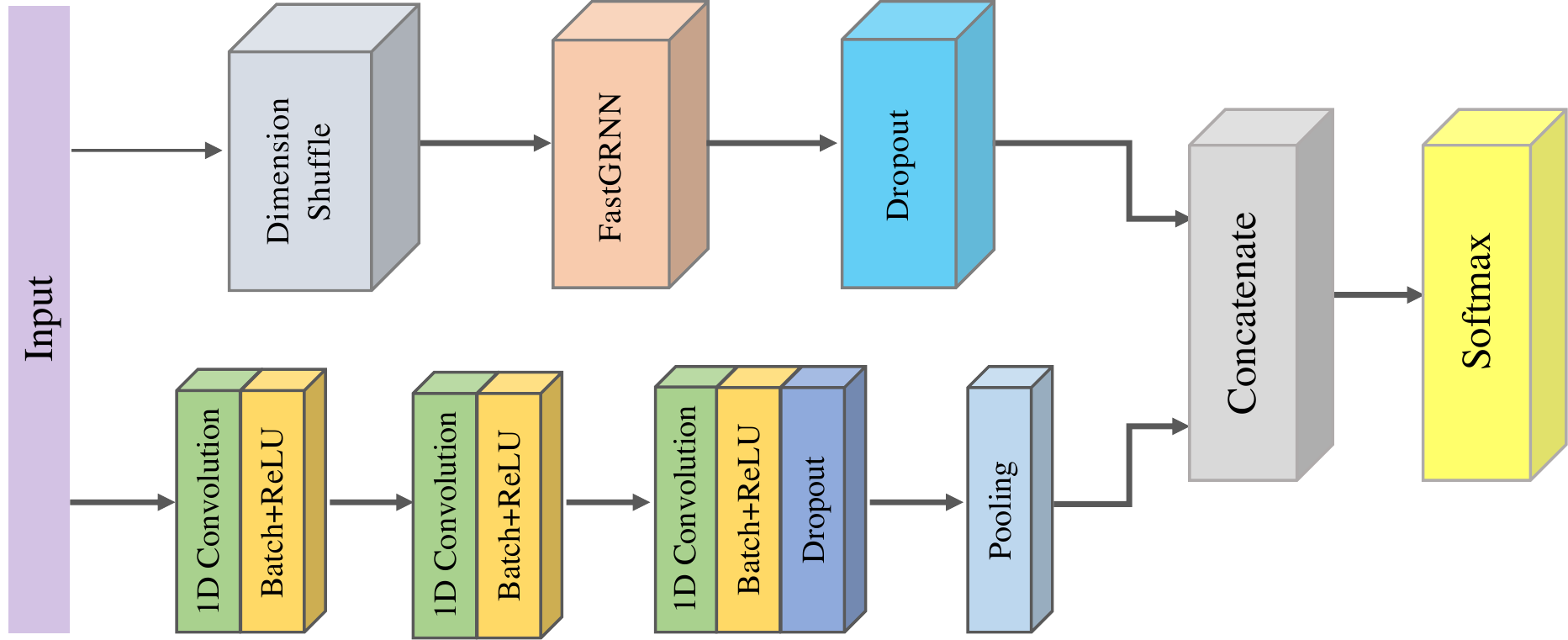}
    \caption{FastGRNN-FCN model architecture}
    \label{fig:fastgrnn_gcn}
\end{figure*}

Data-driven methods can enhance the standard one-dimensional (1D) and two-dimensional (2D) flood propagation models. Recently, a wide range of DL techniques have been used in data-driven flood prediction studies \cite{chang2019flood, sankaranarayanan2019flood}. For example, DL techniques such as conditional Generative Adversarial Neural Networks (cGANs) has been used to enhance the modeling efficiency of the 2D models for urban flood perdition \cite{qian2019physics}. Deep learning neural network has also been adopted to predict the streamflow of Yangtze River \cite{liu2020streamflow}. ML techniques have also been used in hydrologic prediction \cite{fotovatikhah2018survey, gholami2016design}. For example, Ghaderi, Motamedvaziri, Vafakhah, \& Dehghani \cite{ghaderi2019regional} used ML techniques such as support vector machine (SVM) and genetic expression programming (GEP) for flood frequency analysis to improve flood risk management. SVM has also been used to improve flood warning systems by developing a probabilistic rainfall inundation model \cite{pan2019data} and to assess flood risks \cite{opella2019developing}. Moreover, modular local support vectors regression (SVR) models and local artificial neural networks (ANN) models have been employed to predict rainfall level \cite{wu2013prediction}. Ying-Yang firefly algorithm (YYFA) have been used to rainstorm intensity computation for improving urban drainage design \cite{wang2020yin}. A mixed LUBE method enhanced via Multi-Objective Fully Informed Particle Swarm (MOFIPS) have been used for streamflow discharge forecasting \cite{taormina2015ann}. ML methods are also integrated with other techniques such as Internet-of-Things (IoT) applications to predict the flood disasters \cite{mitra2016flood, bande2017smart}. In addition to ML and DL models, other soft computing approaches such as reinforcement learning has been applied for water system control and flood prediction \cite{bhattacharya2003neural}. For example, the use of boosting reinforcement learning algorithm is shown to be effective for flood forecasting \cite{li2016new}. 

The studies discussed above leverage the capabilities of ML and DL for flood prediction and provide decision-making tools for planning for flood risk reduction and emergency response \cite{chen2020modeling, sit2019decentralized}. Due to the recent advancements in sensor technology that enable the gathering of the spatial-temporal status of flood gauges, multivariate time series analysis and flood prediction can be achieved using the state-of-the-art techniques in DL. Moreover, the prediction accuracy can be further be improved when more types of data are considered such as land use and soil characteristics of the sensor location \cite{das2019fb}. However, the adoption of data-driven methods and deep learning for predictive flood warning and situation awareness has been rather limited. Hence, there is a need for the transition in data-driven flood prediction toward applying multivariate time series classification techniques for near real-time flood prediction \cite{khac2012data}. To address this gap, this study proposes a DL model, FastGRNN-FCN, to predict near real-time flood propagation using different data such as rainfall, water levels, geographical information, and the spatial-topological relationship of the network of flood sensors on channel infrastructure.

\section{FastGRNN-FCN Model Architecture}
\label{sec:fastgrnn_fcn}

This study proposes a hybrid model FastGRNN-FCN, as shown in Figure \ref{fig:fastgrnn_gcn} to predict the inundation probability of the flood channel network based on the sensor-collected flooding information. We selected FastGRNN due to its lower training times, lower prediction costs, and higher prediction accuracy \cite{kusupati2018fastgrnn}. The FCN has shown its superior capability in classifying time series \cite{karim2017lstm}. We explain each component of the hybrid model in detail below.

\textbf{FastGRNN:} Parameters of the FastGRNN can be defined by the matrices $\textbf{W} \in \mathbb{R}^{\hat{D}\times D}$, $\textbf{U} \in \mathbb{R}^{\hat{D}\times \hat{D}}$, and bias vector $\textbf{b} \in \mathbb{R}^{\hat{D}}$. $\textbf{X} = [x_1, ..., x_t, ..., x_T]$ is the input data, where $x_t \in \mathbb{R}^D$ represents the $t$-th step feature vector. The goal of multi-class FastGRNN is to learn a function $F: \mathbb{R}^{D\times T} \to {1, ..., L}$ that predicts one of $L$ classes for the given data point $\textbf{X}$. The procedure of the FastGRNN is given by \cite{kusupati2018fastgrnn}:
\begin{align}
\label{eq:fastgrnn}
\textbf{Z}_t &= \sigma(\textbf{W}\textbf{x}_t + \textbf{U}\textbf{h}_{t-1} + \textbf{b}_z) \nonumber\\
\tilde{\textbf{h}}_t &= tanh(\textbf{W}\textbf{x}_t + \textbf{U}\textbf{h}_{t-1} + \textbf{b}_h) \\
\textbf{h}_t &= (\zeta(1-\textbf{z}_t) + \nu) \odot \tilde{\textbf{h}}_t + z_t \odot \textbf{h}_{t-1} \nonumber
\end{align}
where $0 \leq \zeta, \nu \leq 1$ are trainable parameters which are parameterized by the sigmoid function. $\sigma: \mathbb{R} \to \mathbb{R}$ is a non-linear function, such as $tanh$ and sigmoid. $\odot$ denotes the Hadamard product, i.e., $(a \odot b)_i = a_i, b_i$. These parameters of FastGRNN $\Theta_{\text{FastGRNN}} = (\textbf{W}^i, \textbf{U}^i, \textbf{b}_h, \textbf{b}_z, \zeta, \nu)$ are trained jointly using stochastic optimization methods to optimize the loss function $L$ (e.g., softmax cross-entropy) in Eq.(\ref{eq:fastgrnn_op}). 
\begin{align}
\centering
& \min_{\Theta_{\text{FastGRNN}}, \|\textbf{W}^i\|_0 \leq s_{\omega}^{i}, \|\textbf{U}^i\|_{0}\leq s_{u}^{i}, i\in{1,2}} \jmath (\Theta_{\text{FastGRNN}})\nonumber\\
&= \frac{1}{n} \sum_j L(\textbf{X}_j, y_j; \Theta_{\text{FastGRNN}})
\label{eq:fastgrnn_op}
\end{align}

The dimension shuffle prior to FastGRNN can transpose a time series of length $N$ into a multivariate time series of $N$ variables with a single time step. In doing so, an input of $M$ variables and $N$ time steps into a batch of $N$ variables and $M$ time steps. As long as the number of $M$ is significantly smaller than $N$, the training efficiency can be dramatically improved as FastGRNN can process a batch of $N$ variables in $M$ time steps \cite{karim2017lstm}. Following the FastGRNN block, a dropout layer with a ratio of $0.5$ is concatenated. 

\textbf{FCN:} The basic block of an FCN is a convolutional layer followed by a batch normalization (BN) layer and a rectified linear unit (ReLU) activation layer. The batch normalization (momentum of 0.99, epsilon of 0.001) is applied to accelerate the convergence speed and improve generalization. The filter sizes of the three temporal convolutional block are 128, 256, and 128, respectively. To maintain the reliability but control the added randomness of the model, only one dropout layer is included in the model. Following the third FCN block, a dropout layer with a ratio of $0.5$ is used to reduce the feature size and avoid overfitting. After all the three convolutional blocks, the results are then fed into a pooling layer which can significantly reduce the number of parameters before the classification. FCN shows compelling quality and efficiency in testing different datasets \cite{wang2017time}. In addition, FCN does not require heavy data preprocessing or feature engineering \cite{karim2017lstm}. Therefore, we select FCN for the multivariate time series classification in this paper. The basic convolution block is obtained from \cite{wang2017time}:
\begin{align}
    y &= \textbf{W} \otimes \textbf{x} + \textbf{b} \nonumber \\
    s &= BN(y) \\
    h &= ReLU(s) \nonumber
\end{align}
where $\textbf{W}$ is the weight matrix and $\textbf{b}$ is the bias vector. In contrast, the FCN treats a time series of length $N$ as a univariate time series with $N$ time steps. 

Through combining the FastGRNN with FCN, we can classify the time series of collected data using a softmax function. Among other loss functions (e.g., mean squared error loss, exceptional loss, and hinge loss), we selected cross-entropy loss function because it greatly improves the performance of models with softmax outputs, which usually suffer from saturation and slow learning when using the mean squared error loss. The input of the model is a tensor of shape $(N, Q, M)$, where $N$ is the number of samples, $Q$ is the maximum number of time steps amongst all variables, and $M$ is the number of variables in each time step. Each sample has a label representing the class such that a tensor with the shape $(N, 1)$ is the target. The shape of input varies according to different datasets. In the following section, we discuss model training and implementation using the channel sensor data from four flooding events in Harris County, Texas.

\section{Application of the Hybrid Deep Learning Model in Flood Prediction in Harris County}
\label{sec:experiment}

Given the significance of predictive flood warning and situation awareness in urban areas, we employed the proposed FastGRNN-FCN model in the context of Harris County, Texas, which is one of the most flood-prone cities in the U.S. \cite{kapucu2014disaster}. In the context of the flood control channel network, there are $N$ flood sensors' $M$ variables with $Q$ time steps of flooding status data. The input data for both training and testing are in the form of tensor, where the first dimension represents different samples, the second dimension indicates different variables, and the third dimension represents the corresponding value at a particular time interval. Labels are binary values representing inundation status (i.e., if the water level in six hours is higher than the top of the channel or not), and each sample has one label. Feeding the model with the data in the aforementioned tensor format, the future flooding status of the flood sensor area can be predicted. The resulting predictions would provide important insights for flood warning and situation awareness.  
\subsection{Study site}
\begin{figure}[!ht]
    \centering
    \includegraphics[scale=0.36]{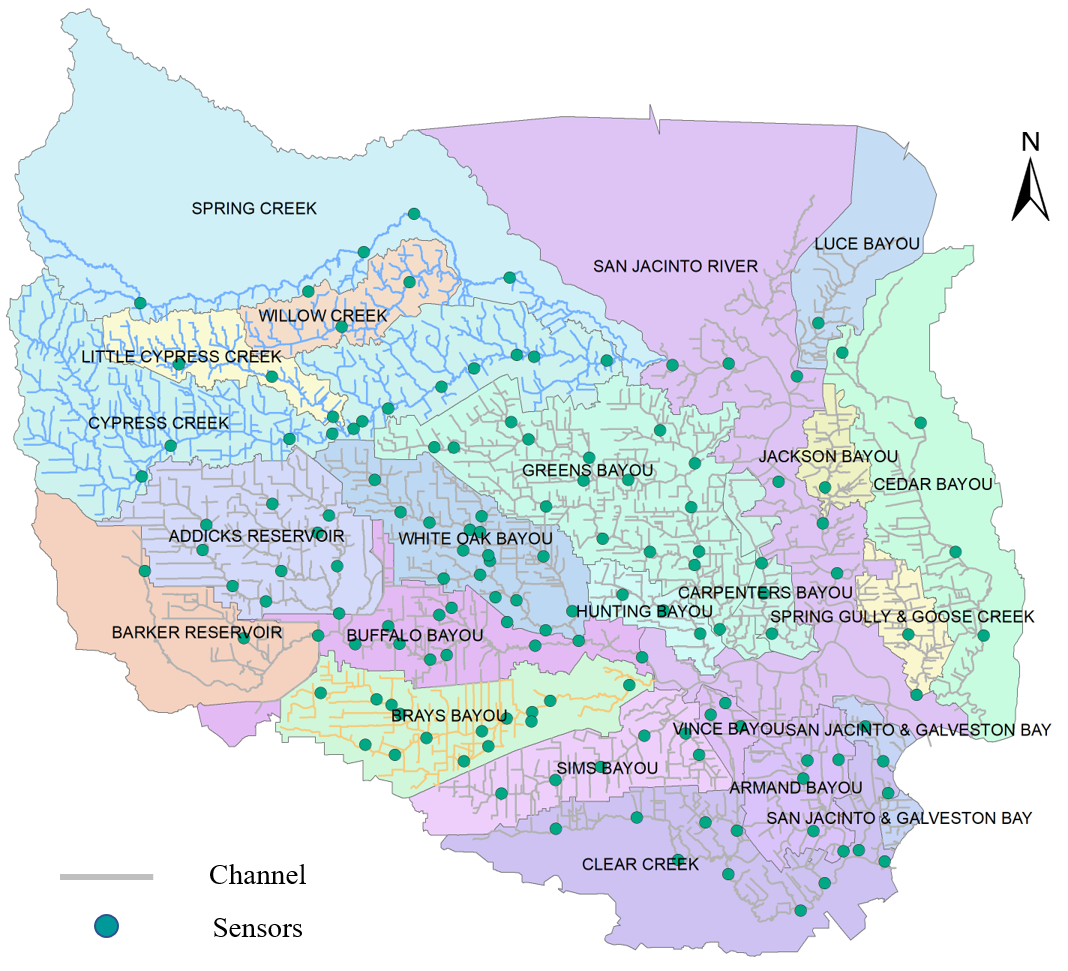}
    \caption{Distribution of flood channel sensors across Harris County}
    \label{fig:harris_county}
\end{figure} 

\begin{table*}[!ht]
\centering
\caption{Input variable summary}
	\begin{tabular}{ll}
		\hline
		\multicolumn{1}{c}{\textbf{Variable}} & \multicolumn{1}{c}{\textbf{Meaning}}                                                                                   \\ \hline
		Rainfall amount                       & Rainfall amount of this sensor during the future 6 hours.                   \\
		Water level in channel                & Current water level of this sensor.                                                                                    \\
		Predecessor rainfall amount           & Average rainfall amount of predecessor sensor(s) \\ during the future 6 hours. \\
		Predecessor water level in channel    & Average current water level of predecessor sensor(s).                                                                  \\
		Successor rainfall amount             & Average rainfall amount of successor sensor(s) during the future 6 hours.   \\
		Successor water level in channel      & Average current water level of successor sensor(s).                                                                    \\
		Impermeable surface percentage      & Average impermeable surface value of all the channels that involves this sensor.                                                                       \\
		Upstream cross-sectional area                     & Summation of the cross-sectional area of all adjacent predecessor sensor(s).                 \\
		Downstream cross-sectional area                   & Summation of the cross-sectional area of all adjacent successor sensor(s).                  \\ \hline
	\end{tabular}
	\label{tab:variables}
\end{table*}

Harris County is one of the most flood-prone urban areas in the U.S. Of 91 weather events that occurred in the U.S. between 2010 and 2017, 43 affected Texas directly \cite{winfree2019need}. Harris County, home to the fourth largest city in the nation-Houston, is the most populated county in Texas and the third most populous county in the United States. Hurricane Harvey, a Category 4 hurricane that hit the Texas coastline on August 25, 2017, is the most significant tropical cyclone rainfall event in the United States history since reliable records began around the 1880s and causing \$190 billion in damage \cite{winfree2019need}. Harris County has more than 2,500 miles (4,023 km) of channels and 22 watersheds that drain the rainfall-runoff and stormwater to a flood control infrastructure and eventually drains into Galveston Bay as suggested by Harris County Flood Control District. \cite{HCFCD2019a}. 

To monitor the flooding in the region, Harris County Flood Control District (HCFCD) deployed 175 flood sensors, as shown in Figure \ref{fig:harris_county}, to track the flooding status across the county at major bayous and creeks and visualized the results on Harris County Flood Warning System \cite{HCFWS2019}. Each sensor collects the water height for the closest channel sections. In addition, each sensor also records the rainfall amount of the location. Combining the flood control network topology, we obtain a repository of spatial-temporal flood data, which can be used for predicting the occurrence of future flooding at different sensor locations. Since Harris County is highly prone to flood and also has sensor data from past flooding events for deep learning model training and implementation, it is an ideal testbed for demonstrating the capability of the proposed FastGRNN-FCN model. The current flood warning system employed lacks predictive elements to enhance proactive response and situation awareness. The application of the FastGRNN-FCN model in the context of Harris County could demonstrate the capabilities and potential of the proposed model for adoption in other regions. 

\subsection{Data processing}

Each sensor is mapped and connected to the corresponding channel. We selected several parameters to train the proposed deep learning model, including weather information such as rainfall amount, hydrological information such as water level in the channel and the channel cross-section area, and the geographic feature such as the impermeable surface percentage. These parameters influence flood propagation in channel networks. Additionally, each channel is not independent. Instead, channels form a network and their connected counterparts also affects a channel section’s capability of discharging rainfall runoff \cite{dong2019b, dong2020c}. Therefore, we also considered each sensor’s successor and predecessor’s information, which help us to embed the network topology into the model training. Eventually, nine categories of data are considered in the proposed model (in Table \ref{tab:variables}) with an example of the dataset is presented in Table \ref{tab:data_sample}.

\begin{table*}[!ht]
\centering
\caption{Example of flood data structure}
\begin{tabular}{lcccccc}
\hline
Sensor                                          & \multicolumn{3}{c}{A100-00-00\_1516}                                                                                               & \multicolumn{3}{c}{D100-00-00\_0524}                                                                                               \\ \hline
Time                                            & \begin{tabular}[c]{@{}c@{}}2019/09/17 \\ 00:00\end{tabular} & \begin{tabular}[c]{@{}c@{}}2019/09/17\\  00:30\end{tabular} & ...... & \begin{tabular}[c]{@{}c@{}}2019/09/17 \\ 00:00\end{tabular} & \begin{tabular}[c]{@{}c@{}}2019/09/17 \\ 00:30\end{tabular} & ...... \\
Water level in channel to bank (ft)             & 18.98                                                       & 18.98                                                       & ...... & 18.7855                                                     & 18.7855                                                     & ...... \\
Rainfall Amount (in)                            & 0                                                           & 0                                                           & ...... & 0                                                           & 0                                                           & ...... \\
Predecessor water level in channel to bank (ft) & -1.01                                                       & -1.01                                                       & ...... & 24.3                                                        & 24.3                                                        & ...... \\
Predecessor rainfall amount (in)                & 0                                                           & 0                                                           & ...... & 0                                                           & 0                                                           & ...... \\
Successor water level in channel to bank (ft)   & 4.08                                                        & 4.19                                                        & ...... & 12.36                                                       & 12.36                                                       & ...... \\
Successor rainfall amount (inch)                & 0                                                           & 0                                                           & ...... & 0.04                                                        & 0.04                                                        & ...... \\
Impermeable surface percentage (\%)             & 15.9                                                        & 15.9                                                        & ...... & 89.55                                                       & 89.55                                                       & ...... \\
Upstream cross-sectional area ($\text{ft}^2$)             & 813.80                                                      & 813.80                                                      & ...... & 4180.60                                                     & 4180.60                                                     & ...... \\
Downstream cross-sectional area ($\text{ft}^2$)           & 1637.30                                                     & 1637.30                                                     & ...... & 6082.31                                                     & 6082.31                                                     & ...... \\ \hline
\end{tabular}
\label{tab:data_sample}
\end{table*}

To train and test the FastGRNN-FCN model, we used data related to four recent flooding events including 2016 Tax Day Flood (4/16/2016 - 4/17/2016), 2016 Memorial Day Flood (5/27/2016 - 5/28/2016), 2017 Hurricane Harvey Flood (8/25/2017 - 8/31/2017), and 2019 Storm Imelda Flood (9/19/2019 - 9/22/2019). Only four flooding events are used in this study. This is because, due to the installation and maintenance cost, the Harris County Flood Control District did not install as many sensors as needed at once, but instead, the sensors have been deployed over the years. As a matter of fact, after Hurricane Harvey, they recognized the importance of a comprehensive understanding of the flood risk. Therefore, more sensors will be installed to have better monitoring of the network flood status and we are hoping to have additional flood sensor data to improve the current model. In this study, to include more sensors and to balance the trade-off between the number of flood events and sensors, we eliminated some flooding events since some of the sensors were newly installed after Tax Day floods, and chose to use the four recent flood events in this study. Additionally, to better inform future sensor installation, the team is analyzing the network observability of existing sensors using network control theory to identify a more effective sensor network configuration. (e.g., number of sensors, location). More flood events and flood sensors could be added to the dataset in the future to enhance the performance of the proposed model. With more flood events to train and validate the model, the robustness and the generalization capacity of the model could be further enhanced. 

We used the data of the first three flood events for training and validating the model, and used Imelda flood data to test the model. The periods of 3-5 days and intervals of 30 minutes are set for each event such that different data about rainfall amount, water level in the channel, and inundation status can be gathered at each time step. The training and validation dataset in the shape of (80997, 9, 96) is obtained and shuffled for training and validating the model. Here, 80997 indicates the number of the data series samples. 9 represents the number of variables considered in the model. The model divides the rainfall amounts and water level in the channel in a two-day interval such that each sample has a sequence length of 96. Similarly, a test dataset in the shape of (29999, 9, 96) is obtained.

\section{Analysis Results }
\label{sec:results}

Using the processed three flood events (i.e., Tax Day flood, Memorial Day flood, and Hurricane Harvey) as inputs, the model is trained, validated, and tested to determine the set of parameters that can achieve the best model performance. In order to reduce the randomness in the model performance characterization, we conducted a Monte-Carlo simulation with 10 repeated runs at each scenarios (i.e., weights) to get the average performance for the model parameter selection. Finally, we tested the model (with the best set of parameters) in prediction of Imelda flooding. 

\subsection{Model performance}
\label{sec:performance}

The constructed 80,997 data instances comprise the dataset for the model training and validation. Early stop is applied to detect the number of epochs needed to avoid the overfitting and increase the training speed. With the prepared multi-variate time series dataset, we tested the performance of the proposed FastGRNN-FCN model. To conduct the experiment, we used a machine with CPU of Intel$\textcircled{\text{R}}$ Core$^{\text{TM}}$ i7-9750H with 2.60 GHz frequency and GPU of NVIDIA GeForce RTX 2060 with 1365 MHz frequency. The training and validation processes take around 500 epochs with each epoch takes around 17 seconds. The model achieves both training and validation accuracy of 0.98 and test accuracy of 0.97. Here, the accuracy is defined as Eq.(\ref{eq:accuracy}):

\begin{equation}
    \text{Accuracy} = \frac{\sum\text{True positive} + \sum\text{True negative}}{\text{Total population}}
    \label{eq:accuracy}
\end{equation}

Several steps were taken to ensure the model yields accurate predictions. First, in this study, the dataset is imbalanced. Therefore, in the case of an imbalanced dataset that contains less positive cases (i.e., flooded) and more negative case (i.e., non-flooded), more weights should be given to the positive case (or minority case), which otherwise would lead to low correct prediction rate as the classifier would favor majority case due to its large number. Using a weighting parameter, the weighted cross entropy loss function can be updated as 
\begin{align}
\centering
& \text{Cross entropy loss} \nonumber\\
&= -(\text{selected weight} \times \text{target} \times log(\text{output})\\
&+ 1.0 \times (1-\text{target}) \times log(1.0-\text{output}))\nonumber
\label{eq:cross_entropy}
\end{align}

Among the many loss functions, we chose the cross entropy loss function becuase (1) each value ranges in [0, 1] and (2) the sum of all values equals to 1, which makes it a nice function to model probability distribution. Essentially, the weight is telling the model to pay more "attention" to samples from an underrepresented class. Using different loss function weights when training the model, the model performance changes accordingly. However, the change of accuracy is minor. As $w$ increase from 1 to 100, the accuracy decreases from 0.978 to 0.976.

However, there remain several issues with the current model. First, considering accuracy to determine the model performance alone can, however, be misleading in certain cases. For example, the accuracy can be dominated by the correct predicted non-flooded cases. In this case, correctly predicted flooded instances make little difference in the final accuracy results, and thus, potentially resulting in incorrect parameter tuning and model selection. Therefore, other means of performance evaluation metrics are needed to select the appropriate model. Second, given the predicted probability, a threshold ($\phi$) is normally used to determine a node's flooding status (a probability greater than the threshold will be considered as flooded). A threshold ($\phi$) of 0.5 is normally used to make the prediction. However, depending on the feature of the dataset (e.g., imbalance rate), there exists a critical threshold ($\phi_c$) that enables the optimal prediction performance. The following subsection discusses the impact of weights on the model performance and the procedure for eventually selecting the optimal sets of parameters for the model. 

\subsubsection{Precision-recall}
\label{sec:weight}

\begin{figure*}[!ht]
    \centering
    \includegraphics[scale=0.35]{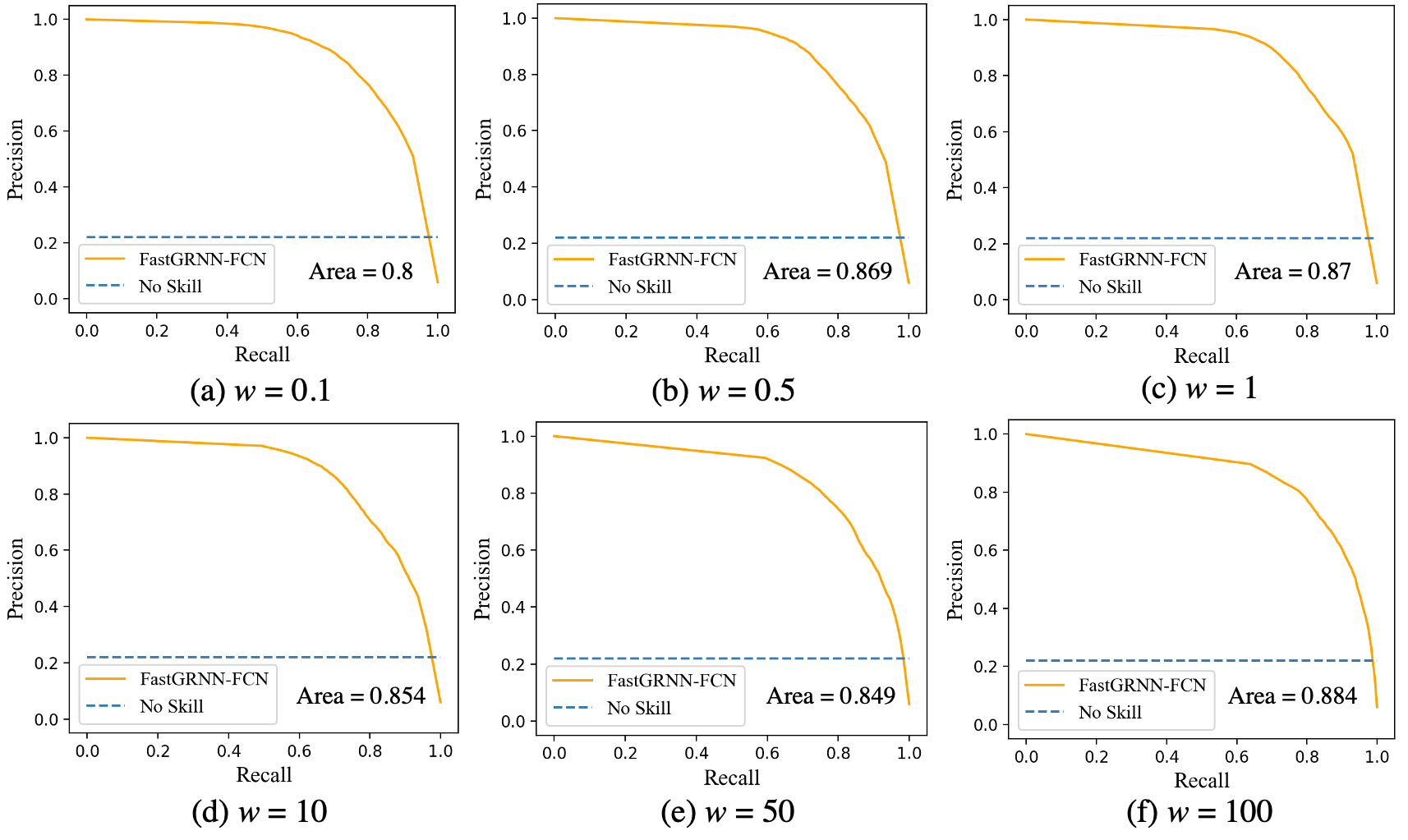}
    \caption{Precision-recall curve comparison at different weights. Blue dash line shows the prediction performance when a naive classification model that treats every data point as flooded. The orange solid line represents the FastGRNN-FCN prediction performance. The area underneath the orange solid curve is also an indicator of the FastGRNN-FCN performance, with higher value representing a better performance. }
    \label{fig:PR_curve}
\end{figure*} 

In addition to the accuracy metric, confusion matrix can provide insights into details of the prediction results, such as which classes are being predicted correctly or incorrectly, and what type of errors are being made. The hypothesis is that the studied node is flooded, true positive (TP) indicates the model correctly identifies flood case and false negative (FN) indicates the model incorrectly rejects the hypothesis. Similarly, false positive (FP) denotes the model incorrectly identifies flooding case and true negative (TN) denotes the model correctly rejects the hypothesis. 

With the confusion matrix, we can derive the precision metric (TP / (TP + FP)) that can be used to quantify the number of correct positive predictions made, and recall metric (TP / (TP + FN)) that measures the number of correct positive predictions made out of all positive predictions that could have been made. Precision ranges between 0.0 (no precision) and 1.0 (perfect precision), while recall ranges between 0.0 (no recall) and 1.0 (full recall). Both the precision and the recall mainly focus on the minority cases (i.e., flooded nodes) and are unconcerned with the majority class (i.e., non-flooded nodes). Therefore, Instead of the commonly used Receiver Operating Characteristic (ROC), which usually yields an excessively optimistic view of model performance, we adopted precision and recall measures that enable the assessment of the performance for a classifier based on the minority class when data is imbalanced \cite{he2013imbalanced}. A well-performing model would have a curve that bends towards the coordinate of (1,1), while a no-skill classifier would yield a horizontal line with a precision that is proportional to the number of positive examples in the dataset, which is 1:3.5 (i.e., positive cases: negative cases) in our dataset. For a balanced dataset, the no-skill line would be at precision = 0.5. With the precision and recall derived from the confusion matrix, we obtained the precision-recall curve by plotting the precision on the y-axis and the recall on the x-axis for different probability thresholds ($\phi$). 

In the curve shown in Figure \ref{fig:PR_curve}, the yellow solid line shows the predicting capability of the FastGRNN-FCN model and the blue dash line describes a method that has no skills (i.e., no predicting capability). With various weights employed during the model training, FastGRNN-FCN presents different prediction performances. To quantitatively measure the model performance, we calculated the area under the curve and use it to represent the prediction capability of the proposed FastGRNN-FCN model. As we can conclude from the figure, FastGRNN-FCN performs the best at the weight of 1 with an area under curve of 0.87 and decreases as the weight increases.

\subsubsection{F-measure}

\begin{figure*}[!ht]
    \centering
    \includegraphics[scale=0.355]{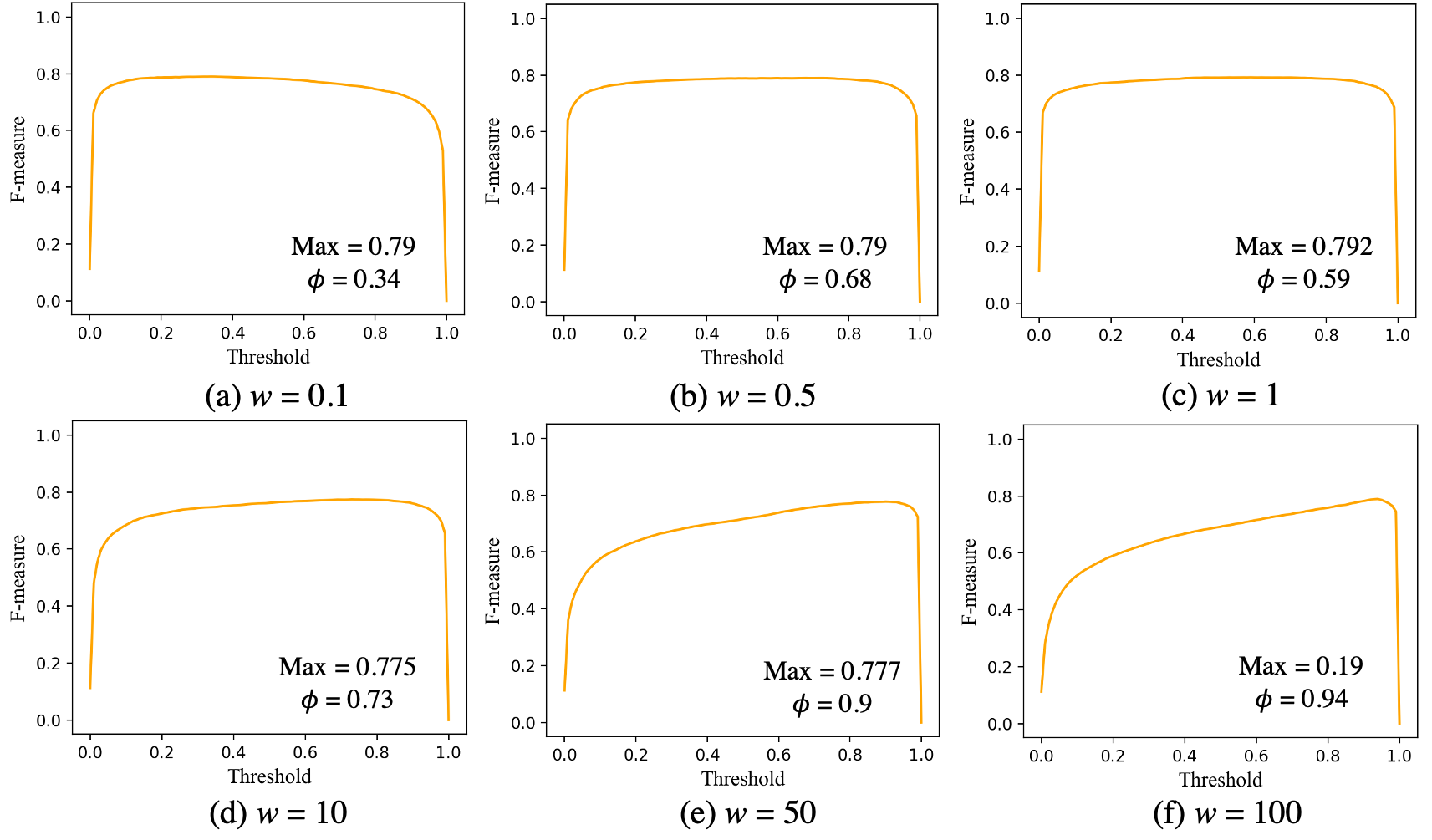}
    \caption{F-measure comparison at different weights. At each weight, a maximum F-measure can be found and the corresponding threshold that enables the maximum F-measure is identified as the critical threshold.}
    \label{fig:F_measure}
\end{figure*}

For prediction tasks using an imbalanced dataset, the objective is to improve recall without hurting precision. These goals are, however, conflicting with each other. For example, to increase the TP for the minority class, the number of FP will also be increased, leading to reduced precision. That is to say, neither precision nor recall can fully capture the model prediction performance. We can have great precision with low recall, or vice versa. Therefore, similar to the widely adopted classification accuracy in which only a single measure is used to summarize model performance, we employed the F-measure, which combines both the precision and recall into one single score that encapsulates both properties. Through training the model with different parameters (e.g., weight $w$ and threshold $\phi$) with multiple runs, we seek to find the optimal set of parameters that enables the highest F-measure. The F-measure is calculated as shown in Equation (\ref{eq:f_measure}). 

\begin{equation}
    \text{F-Measure} = 2 \times \frac{\text{Precision}\times \text{Recall}}{\text{Precision}+\text{Recall}}
\label{eq:f_measure}
\end{equation}

In addition, the threshold ($\phi$) used to classify the prediction flooding probability is also critical in determining model prediction performance. A neutral value of 0.5 indicates the model has no preference between flooded and non-flooded cases, which can be less efficient in the case of an imbalanced dataset. Therefore, in this study, we also tested the model prediction performance using different threshold values ($\phi$) to identify the suitable critical threshold ($\phi_c$) for flooding classification. By examining the calculated F-measure for different selected weights examples in Figure \ref{fig:F_measure}, we can see that the proposed FastGRNN-FCN model shows its best performance (among the six examples) in predicting flood at the weight of 1, by achieving the maximum F-measure of 0.792 at the critical threshold ($\phi_c$) of 0.59. Additionally, as the weight increases, the critical threshold required for achieving the maximum F-measure also increases.

\subsection{Monte-Carlo Simulation}
\label{sec:monte_carlo}

\begin{figure}[!ht]
    \centering
    \includegraphics[scale=0.36]{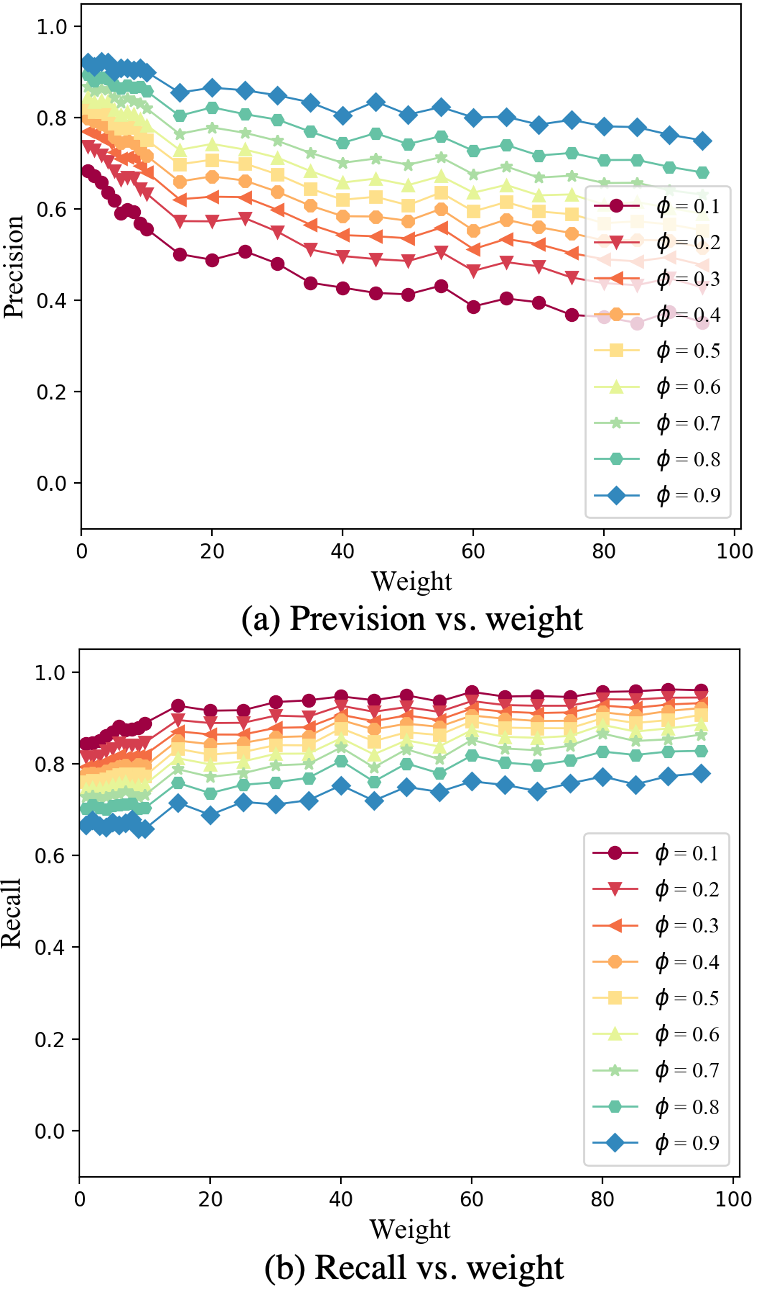}
    \caption{Impact of weights on precision and recall at different thresholds. Higher threshold results in higher precision but lower recall. Higher weight leads to lower precision but higher recall.}
    \label{fig:precision_recall_weight_w_different_threshold}
\end{figure}

From the examples above, we can see that model performance can be influenced by both weight and critical threshold. In order to identify the optimal set of weight and critical threshold that enables the best prediction performance, a more comprehensive analysis (instead of six scenarios) is needed. Therefore, we conducted experiments on weights between 1 and 100, with step of 1 in [1, 10] (e.g., 1, 2, 3, ..., 10) and step of 5 in [10, 100] (e.g., 10, 15, 20, ..., 100). 

\begin{table}[!ht]
\centering
\caption{Model performance characterization at different weights}
\begin{tabular}{cccc}
\hline
Weight & \begin{tabular}[c]{@{}c@{}}Max. \\ F-measure\end{tabular} & \begin{tabular}[c]{@{}c@{}}F-measure \\ Curve Area\end{tabular} & \begin{tabular}[c]{@{}c@{}}Precision-Recall \\ Curve Area\end{tabular} \\ \hline
0.1    & 0.79                                                      & 0.752                                                           & 0.871                                                                  \\
0.5    & 0.79                                                      & 0.765                                                           & 0.869                                                                  \\
1      & 0.792                                                     & 0.768                                                           & 0.87                                                                   \\
2      & 0.791                                                     & 0.766                                                           & 0.862                                                                  \\
3      & 0.791                                                     & 0.762                                                           & 0.863                                                                  \\
4      & 0.791                                                     & 0.759                                                           & 0.868                                                                  \\
5      & 0.785                                                     & 0.754                                                           & 0.864                                                                  \\
6      & 0.782                                                     & 0.747                                                           & 0.86                                                                   \\
7      & 0.788                                                     & 0.751                                                           & 0.864                                                                  \\
8      & 0.785                                                     & 0.748                                                           & 0.861                                                                  \\
9      & 0.777                                                     & 0.738                                                           & 0.86                                                                   \\
10     & 0.775                                                     & 0.732                                                           & 0.854                                                                  \\
15     & 0.783                                                     & 0.726                                                           & 0.858                                                                  \\
20     & 0.779                                                     & 0.721                                                           & 0.85                                                                   \\
25     & 0.783                                                     & 0.727                                                           & 0.859                                                                  \\
30     & 0.778                                                     & 0.715                                                           & 0.856                                                                  \\
35     & 0.774                                                     & 0.697                                                           & 0.849                                                                  \\
40     & 0.779                                                     & 0.695                                                           & 0.854                                                                  \\
45     & 0.774                                                     & 0.686                                                           & 0.846                                                                  \\
50     & 0.777                                                     & 0.688                                                           & 0.849                                                                  \\
55     & 0.78                                                      & 0.698                                                           & 0.853                                                                  \\
60     & 0.783                                                     & 0.678                                                           & 0.852                                                                  \\
65     & 0.78                                                      & 0.688                                                           & 0.848                                                                  \\
70     & 0.765                                                     & 0.675                                                           & 0.838                                                                  \\
75     & 0.776                                                     & 0.669                                                           & 0.848                                                                  \\
80     & 0.779                                                     & 0.666                                                           & 0.85                                                                   \\
85     & 0.768                                                     & 0.658                                                           & 0.838                                                                  \\
90     & 0.77                                                      & 0.663                                                           & 0.845                                                                  \\
95     & 0.765                                                     & 0.654                                                           & 0.837                                                                  \\
100    & 0.79                                                      & 0.661                                                           & 0.853                                                                  \\ \hline
\end{tabular}
\label{tab:model_performance}
\end{table}

\begin{table*}[!ht]
\centering
\caption{Model performance comparison}
\begin{tabular}{lccccc}
\hline
                            & FastGRNN-FCN & ALSTM-FCN & FastGRNN & ALSTM & FCN   \\ \hline
Max. Accuracy               & 0.978        & 0.972     & 0.976    & 0.958 & 0.973 \\
Max. F-Measure              & 0.793        & 0.727     & 0.787    & 0.578 & 0.759 \\
F-Measure Curve Area        & 0.77         & 0.694     & 0.752    & 0.425 & 0.734 \\
Precision-Recall Curve Area & 0.874        & 0.8       & 0.865    & 0.574 & 0.81  \\
Training Time per Epoch (s) & 17           & 25        & 11       & 19    & 12    \\ \hline
\end{tabular}
\label{tab:model_comparison}
\end{table*}

Moreover, due to the randomness in each simulation, the model performance fluctuates in different runs. The randomness comes from the dropout block in the convolutional layer shown in Figure \ref{fig:fastgrnn_gcn}, where the dropout layer randomly disregards 50\% the information and used the rest half to train the model, in order to avoid the overfitting and prevent the model stuck in a local minimum. To eliminate the randomness in the performance evaluation, we employed the Monte-Carlo simulation scheme. Ten runs of the model were conducted at each weight and their average value was calculated. The average performance of multiple runs provides a reliable indicator for the model performance. First, we examined the impact of weights on precision and recall. Figure \ref{fig:precision_recall_weight_w_different_threshold} shows the impact of the weights on precision and recall at different thresholds. In general, as the weight increases, the precision decreases and recall increases. That is because, with the increase of weight, false-positive rate increases and the false-negative rate decreases, which leads to an increase of recall but a decrease of precision. Therefore, there is a trade-off between precision and recall. Hence, we used the area under the precision-recall curve to determine model performance. As shown in Table \ref{tab:model_performance}, the area underneath the precision-recall curve achieves the highest value at the weight of 1.  

Since the F-measure has shown its capability in providing a balanced criterion for selecting the optimal model and model parameters, we also examined the maximum F-measure at different thresholds. F-measure is derived from precision and recall, which are related to the TP, FP, and FN rate. With the increase of weight, TP rate and FP rate increase, and FN rate decreases, which leads to the growth of recall but the decline of precision. Therefore, there should be a weight that reaches a balance point for both precision and recall and give the best F-measure. Table \ref{tab:model_performance} shows the maximum F-measure and the area under the precision-recall curve at each weight. The results show that the model has the highest maximum F-measure at the weight of 1. According to the analysis in Figure \ref{fig:F_measure}, we concluded that the corresponding critical threshold ($\phi_c$) of the proposed FastGRNN-FCN model to achieve the best flood prediction is at 0.59. To show the prediction capability of the proposed FastGRNN-FCN model, we tested other state-of-art models and presented the performance comparison in Table \ref{tab:model_comparison}. The results show a superior performance of FastGRNN-FCN model on flood prediction.

\subsection{Flood prediction of storm Imelda}
\label{sec:imelda}

\begin{figure*}[!ht]
    \centering
    \includegraphics[scale=0.28]{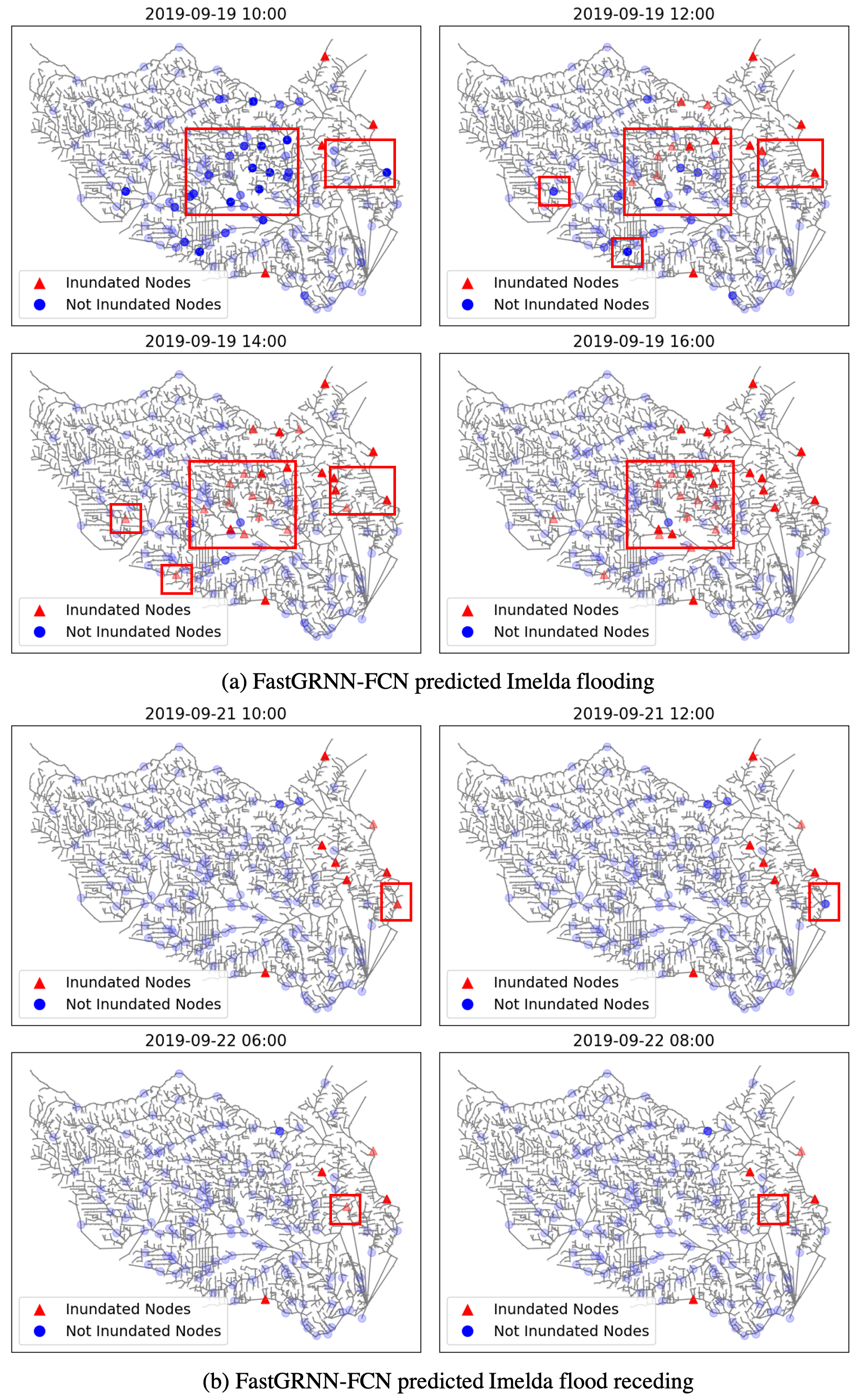}
    \caption{FastGRNN-FCN predicted Imelda flooding and receding. The darkness of the blue describes the predicted flooding probability, with a darker color represents the higher probability. The lightness of the red color describes the inundation of the nodes, with the lighter red color representing the sensor is going to turn into non-inundated, i.e., flood recession.}
    \label{fig:imelda_flooding}
\end{figure*}

Using the trained FastGRNN-FCN model with the weight of 1 and the critical threshold of 0.59, we implemented prediction for the 2019 Imelda flood. We predicted the flood at each flood sensors in a two-hour interval (Figure \ref{fig:imelda_flooding}(a)), with red triangles representing the inundated sensors and blue nodes indicating the non-inundated sensors. 

\begin{figure}[!ht]
    \centering
    \includegraphics[scale=0.35]{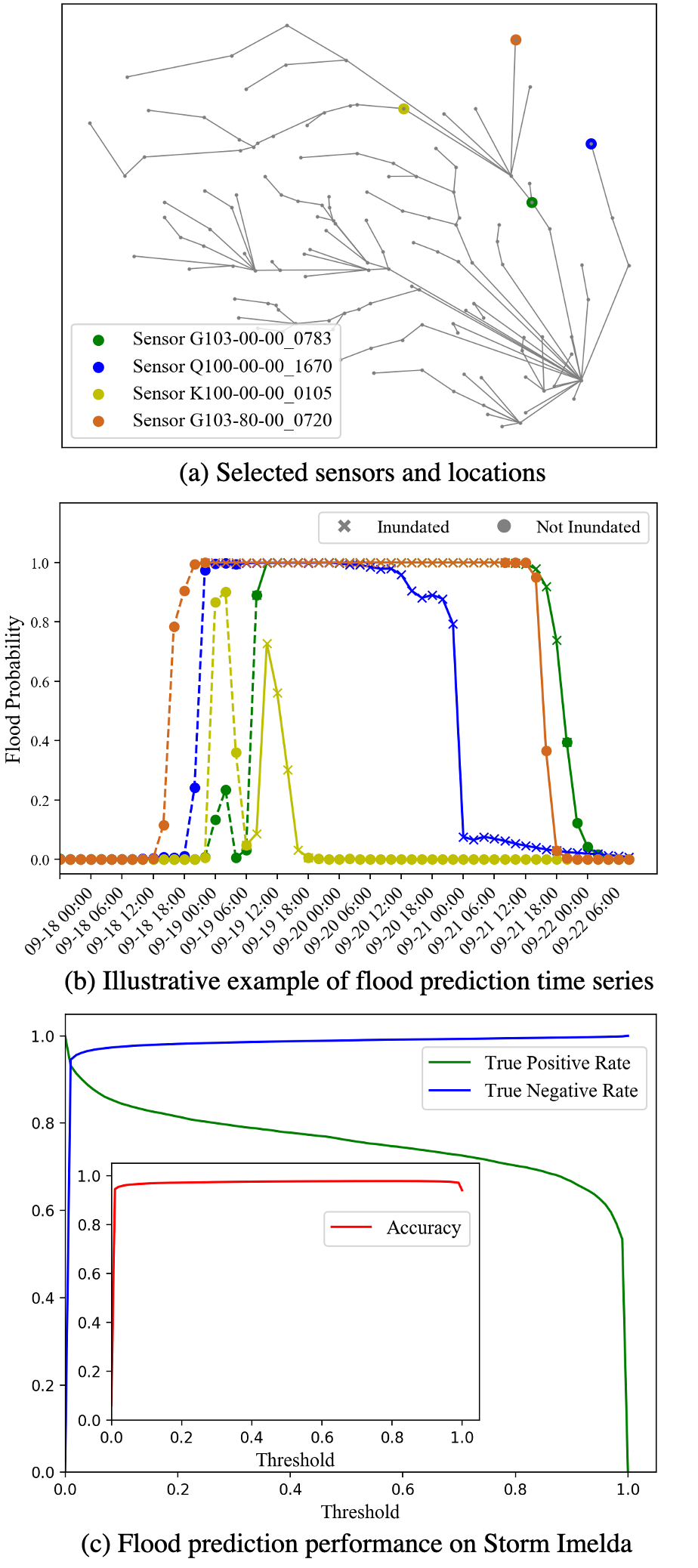}
    \caption{Illustrative time series example of flood prediction}
    \label{fig:flood_time_series}
\end{figure}

Figure \ref{fig:imelda_flooding}(a) shows the predicted flooding and empirical flood spreading process from 2019-09-10 10:00 to 2019-09-19 16:00. By examining the red box highlighted area, we can see that dark blue nodes (sensors with high flooding probability) turned into red triangles in two hours. In addition, the proposed FastGRNN-FCN model is able to predict the flood recession as well. Figure \ref{fig:imelda_flooding}(b) shows an example of the captured flood recession process. As shown in the red box highlighted region, the red triangle turns into a blue dot in two hours. The above two sets of results show the capability of the proposed FastGRNN-FCN model in both predicting the flood propagation, as well as flood recession in a large network. It should be noted that the highlighted nodes are just an illustrative case of the prediction results. If we focus on the nodes outside the box, they only gets darker (or lighter) (i.e., increase or decrease of predicted flooding/receding probability) in 2 hours without turning into red (or blue). These nodes will probably be flooded in next 4 hours or 6 hours and some may never experience flooding, which is the issue of model prediction accuracy. To better show the flood prediction process, Figure \ref{fig:flood_time_series}(b) shows the time series of flood prediction at selected sensors. As we can see, the predicted flood probability increases as the storm Imelda unfolds. Before a sensor status changes to flooded (when the node turns into cross), a high flood probability is predicted by the model. For example, sensor G103 (orange) and Q100 (blue) shows steady prediction on flooding. There also exist fluctuations in the flood prediction. However, both sensors have shown high probability of flooding prior to the actual flooding despite the probability drop. For example, sensor G103 (green) and K100 (yellow) first show a high flooding prediction and the probability then suddenly dropped before flooding. As the rainfall dissipates, the flood probability decreases which indicates flood water receding (cross sign eventually turns into node). Figure \ref{fig:flood_time_series}(c) shows that there is a critical threshold where we can have a balance between TP and TN. Although the model shows a high accuracy with very low threshold, in case of imbalanced dataset (i.e., more unflooded nodes than flooded nodes), accuracy can be dominated by the TN (as in Eq.(\ref{eq:accuracy})). We adopted F-measure and Precision-Recall Curve (Figure \ref{fig:PR_curve} and \ref{fig:F_measure}) as the performance evaluation criteria.

\section{Discussion}
\label{sec:discussion}

The proposed FastGRNN-FCN model shows a high prediction performance with weight and the critical threshold at 1 and 0.59, respectively. Despite “more attention” is given to the minority cases (i.e., flooded instances), a weight of 1 still shows the best performance. This result is mainly due to the data imbalance in the test case (i.e., storm Imelda flooding). The weight in the cross entropy loss function determines the emphasis that the model puts on each case (i.e., positive or negative). A weight of 1 means the model treats both cases the same. However, in this case, the dataset in very imbalanced. For example, the positive to negative ratio for the training and validation sets (i.e., 2016 Tax Day Flood, 2016 Memorial Day Flood, and 2017 Hurricane Harvey Flood) is 1: 3.5. The test set is even more imbalanced with a ratio of 1: 15.77, which is around 4.5 times more than the training and validation set. Due to the data imbalanced issue, even with a weight of 1, the model is still favoring the positive cases, which aligns with our initial goal of giving more attention to the minority cases. Despite the data imbalance issue, the proposed FastGRNN-FCN model still achieves an accuracy of 97.8\% and F-measure of 0.792, which shows the capability of the proposed model for predictive flood warning and situation awareness.

This model is developed to provide early warning for potential flooding sites (i.e., channel overflow) in the next few hours based on the sensor data related to real-time rainfall and water levels. Using the output of the model, potential users only need to prepare the data in the required format (i.e., [x, 9, 96] where x represents the size of data, 9 represents the nine parameters selected, and 96 represents two days data with 30 minutes interval) and then feed the data into the model. The users can choose to use pre-trained model for flood prediction which only takes a couple of minutes to implement, or iteratively train the model when new data is obtained (which can take a couple of hours). The outputs of this model can help office of emergency management in identifying potential areas in risk and alert the communities to take protective actions. Flood control infrastructure operators can also use this information to perform analysis to prioritize the infrastructure protection projects (such as widening channel) to mitigate future flood risk.

\section{Concluding Remarks}
\label{sec:conclusion}

This study presents a hybrid deep learning model, FastGRNN-FCN, and its testing in a case study of flood prediction in Harris County, Texas. The model achieves a prediction accuracy of 97.8\%. Due to the imbalanced nature of the dataset (i.e., less flooded nodes than the non-flooded ones), we employed the precision-recall curve and F-measure to identify the best set of parameters. Through performing tests on 30 weight parameters, we conclude that the model achieves optimal flood prediction performance at a weight of 1 and a critical threshold of 0.59. Using the trained model, we predict the propagation and recession of the 2019 Imelda flooding in two-hour time stamps. The model is not only able to accurately predict the flood cascading process but also captures the flood receding process. This is largely due to the fact that the flood control network structure is embedded in the input data (i.e., the nine variables included in Table \ref{tab:variables}). The proposed FastGRNN-FCN model enables accurate spatial-temporal prediction on flood propagation in channel networks, which provides an effective tool for predictive flood warning and situation awareness to support emergency response decision-making. With this tool, we will be able to automate the flood prediction process by creating a pipeline to constantly collect the stream data from flood gauges and rainfall fall data. The flood data will be sliced into a multivariate time series and then feed into the model. A dashboard can be created to visualize the results to detect the area in high-risk for the next few hours. This can help flood managers and emergency managers to effectively communicate and coordinate with each other to make decisions regarding flood control strategies accordingly such as flood prevention strategies and timely evacuation warning \cite{li2019modeling, dong2020institutinoal}.

The existing flood warning system mainly focuses on monitoring flood propagation. With the growing available infrastructure data and advanced methodology, a computer-aided infrastructure failure prediction model, FastGRNN-FCN, is made possible. The proposed hybrid deep learning model specifically focuses on accurately predicting flood propagation. The case study in Harris County demonstrates the capability of the proposed FastGRNN-FCN model in providing early warning for the occurrence of sequential disruption and cascading failure. Moreover, the application of the proposed FastGRNN-FCN model is not limited to flood prediction, but also suitable for other infrastructure networks, such as power grid and transportation network, and other disciplines, such as biology and information science. Using the records of the components' failure (e.g., congestion, and flow disruption) over time, the cascading failure of the network and the access to critical facilities \cite{dong2019robust} can be then predicted. 

Although the proposed FastGRNN-FCN model shows good performance in predicting the flood cascade, the model can still benefit from improvement in several aspects. First, three flood events are used in this paper to train the FastGRNN-FCN model. However, the advantage of deep learning methods can be further revealed with a larger training dataset. Therefore, with more flood events recorded in the future, the training set can be expanded to provide a more robust and accurate flood prediction. Second, built-environment also plays a critical part in urban flooding management. Knowing the nearby drainage system such as sewer channel can greatly help us to model rainfall runoff \cite{dong2020b}. With the infrastructure interdependencies considered in this model, a more realistic urban flood prediction model can be obtained. Third, deep learning models are advancing rapidly and newer algorithms for classification and time series analysis will be continuingly developed. Further research can benefit from integrating these newly developed techniques and algorithms such as Enhanced Probabilistic Neural Network (EPNN) \cite{ahmadlou2010enhanced} and Neural Dynamic Classification (NDC) \cite{rafiei2017new}. Finally, aside from physical infrastructure data, social sensing data (e.g., crowd rescue, social media, and emergency call) can also capture the flooding status over the network. 

\section*{Acknowledgements} 
    
The authors would like to acknowledge funding support from the National Science Foundation RAPID project \#1760258 and CRISP 2.0 Type 2 \#1832662. Any opinions, conclusion, and recommendations expressed in this research are those of the authors and do not necessarily reflect the view of the funding agencies. The authors would also like to thank the Editor and the anonymous reviewers for their constructive comments and valuable insights to improve the quality of the article. 

\bibliographystyle{unsrt}  
\bibliography{ms}  


\end{document}